\documentstyle[twoside,fleqn,espcrc2,epsf]{article}
\begin{document}

\begin{flushright}
LPTENS-96/41 \\
{\tt hep-th@xxx/9606192}\\
June 1996
\end{flushright}
\begin{center}
\vspace{3 ex}
{\Large\bf Discontinuous BPS spectra}\\
\vspace{1 ex}
{\Large\bf in $N=2$ susy QCD}\\
\vspace{8 ex}
Adel Bilal \\
{\it CNRS - Laboratoire de Physique Th\'eorique de l'\'Ecole
        Normale Sup\'erieure}\\
{\it 24 rue Lhomond, 75231 Paris Cedex 05, France} \\
 {\tt bilal@physique.ens.fr}\\
\vspace{15 ex}
\bf Abstract\\
\end{center}
\vspace{2 ex}
These notes are based on work done in collaboration with Frank Ferrari. 
We show how to determine the spectra of stable BPS states in $N=2$ 
supersymmetric $SU(2)$ Yang-Mills theories that are asymptotically free, 
i.e. without and with $N_f=1,2,3$ quark hypermultiplets. In all cases: \\
$\bullet$ There is a curve of marginal stability diffeomorphic to a circle 
and going through all (finite) singular points of moduli space.  \\
$\bullet$ The BPS spectra are discontinuous across these curves.  \\
$\bullet$ The strong-coupling spectra (inside the curves) contain only 
those BPS states that can become massless and are responsible for the 
singularities. Except for $N_f=3$, they form a multiplet (with different 
masses) of the broken global discrete symmetry.   \\
$\bullet$ All other semi-classical BPS states must and do decay 
consistently when crossing the curves. \\
$\bullet$ The weak-coupling, i.e. semi-classical BPS spectra, contain 
no magnetic charges larger than one for $N_f=0,1,2$ and no magnetic charges 
larger than two for $N_f=3$.

\vspace{15 ex}

\centerline{{\it Talk given at SUSY'96, College Park, MD, June 1996 }}

\thispagestyle{empty}
\newpage

\newcommand{\F}{{\cal F}}
\newcommand{\f}{\phi}
\newcommand{\Z}{{\bf Z}}
\newcommand{\R}{{\bf R}}
\newcommand{\C}{{$\cal C$}}
\newcommand{\sw}{{\cal S}_W}
\newcommand{\ssp}{{\cal S}_{S+}}
\newcommand{\ssm}{{\cal S}_{S-}}
\newcommand{\sso}{{\cal S}_{S0}}
\newcommand{\rw}{{\cal R}_W}
\newcommand{\rsm}{{\cal R}_{S-}}
\newcommand{\rsp}{{\cal R}_{S+}}
\newcommand{\rso}{{\cal R}_{S0}}
\newcommand{\fig}[3]{\epsfxsize=#1\epsfysize=#2\epsfbox{#3}}

 \def\PL #1 #2 #3 {Phys.~Lett.~{\bf #1} (#2) #3}
 \def\NP #1 #2 #3 {Nucl.~Phys.~{\bf #1} (#2) #3}
 \def\PR #1 #2 #3 {Phys.~Rev.~{\bf #1} (#2) #3}
 \def\PRL #1 #2 #3 {Phys.~Rev.~Lett.~{\bf #1} (#2) #3}
 \def\MPL #1 #2 #3 {Mod.~Phys. Lett.~{\bf #1} (#2) #3}
 \def\CMP #1 #2 #3 {Commun. Math. Phys.~{\bf #1} (#2) #3}

\newcommand{\ttbs}{\char'134}
\newcommand{\AmS}{{\protect\the\textfont2
  A\kern-.1667em\lower.5ex\hbox{M}\kern-.125emS}}

\hyphenation{author another created financial paper re-commend-ed}


\title{Discontinuous BPS spectra in $N=2$ susy QCD}

\author{Adel Bilal\address{CNRS - Laboratoire de Physique Th\'eorique de l'\'Ecole
        Normale Sup\'erieure\\ 
        24 rue Lhomond, 75231 Paris Cedex 05, France\\
        {\tt bilal@physique.ens.fr}}
	}

\begin{abstract}
These notes are based on work done in collaboration with Frank Ferrari. 
We show how to determine the spectra of stable BPS states in $N=2$ 
supersymmetric $SU(2)$ Yang-Mills theories that are asymptotically free, 
i.e. without and with $N_f=1,2,3$ quark hypermultiplets. In all cases: \\
$\bullet$ There is a curve of marginal stability diffeomorphic to a circle 
and going through all (finite) singular points of moduli space.  \\
$\bullet$ The BPS spectra are discontinuous across these curves.  \\
$\bullet$ The strong-coupling spectra (inside the curves) contain only 
those BPS states that can become massless and are responsible for the 
singularities. Except for $N_f=3$, they form a multiplet (with different 
masses) of the broken global discrete symmetry.   \\
$\bullet$ All other semi-classical BPS states must and do decay 
consistently when crossing the curves. \\
$\bullet$ The weak-coupling, i.e. semi-classical BPS spectra, contain 
no magnetic charges larger than one for $N_f=0,1,2$ and no magnetic charges 
larger than two for $N_f=3$.
\end{abstract}

\maketitle

\section{INTRODUCTION}

$N=2$ susy gauge theories possess an important holomorphicity property. The low-energy
effective action of the $SU(2)$ theory e.g. is characterized by a
holomorphic function $\F(a)$, or actually by two such functions $a(u)$ and
$a_D(u)=\F'(a(u))$. Knowledge of these two functions completely determines the
low-energy effective action, as well as the masses of all BPS states at any point $u$ of
the moduli space. Holomorphicity is of course a very strong mathematical condition. It
allowed Seiberg and Witten \cite{SW}, given some physical one-loop input, to
completely determine these two functions, thus obtaining also all non-perturbative
multi-instanton contributions. 

One might expect that holomorphicity implies that all physical quantities vary
smoothly over the moduli space. This is certainly true for the masses. There is
however one important exception that concerns the stability of BPS states. As first
observed in a two-dimensional context \cite{CV} and in the present context by Seiberg
and Witten \cite{SW}, there exists a curve on moduli space, called the curve of
marginal stability, where otherwise stable BPS states become degenerate with other BPS
states and can decay. This curve is simply determined by the ratio $(a_D/a)(u)$ being
real, and hence has a priori nothing to do with the holomorphicity properties of
$a_D(u)$ and $a(u)$. In \cite{FB,BF} we studied the properties of these curves, which
together with a certain discrete global symmetry acting on the (Coulomb branch of)
moduli space, allowed us to exactly determine which BPS states can and do exist on
either side of the curves, i.e. the weak-coupling (semi-classical) and strong-coupling 
spectra. These spectra are highly discontinuous across the curves, with almost all 
weak-coupling states decaying into very few strong-coupling states.

Here, after giving a very brief account of the work of Seiberg and Witten for gauge
group $SU(2)$, we discuss the curves of marginal stability 
together with the discrete global symmetries and show how the
different spectra are obtained. We first concentrate in more detail on the pure
$SU(2)$ Yang-Mills case ($N_f=0$), and then present the results for the cases with
quark hypermultiplets, $N_f=1,2,3$, which are the only asymptotically free
possibilities for $SU(2)$ (restricting ourselves to vanishing bare masses).

\section{REVIEW OF SEIBERG-WITTEN THEORY}

Let us start with a flash review of \cite{SW} where the low-energy effective action
and BPS mass formula on the moduli space of pure $N=2$ susy $SU(2)$ Yang-Mills theory
was studied. The relevant susy multiplet is a $N=2$ vector multiplet containing a
vector $A_\mu\equiv \sum_{b=1}^3 A_\mu^b\, {1\over 2} \sigma_b$, 
Weyl spinors $\lambda,\ \psi$ and a complex scalar $\f$, all
in the adjoint representation of the gauge group $SU(2)$. The action is such that a
scalar potential $V(\f)={1\over 2} {\rm tr \, } ([\f, \f^+])^2$ is present. Unbroken
susy implies $[\f, \f^+]=0$, still leaving the possibility of a non-vanishing vavuum
expectation value of $\f$ which one may take as $\f={1\over 2} a\, \sigma_3$
with $a\in {\bf C}$. Gauge
inequivalent vacua then are parametrized by $a^2$ or rather by 
$u=\langle {\rm tr \, } \f^2\rangle$
which is a good coordinate on the moduli space ${\cal M} ={\bf C} \cup \{ \infty \}$.
Since $\f$ has a non-vanishing expectation value, by the Higgs mechanism, all
components $b=1,2$ of the vector multiplet acquire a mass $m=\sqrt{2} \vert a \vert$
while the $b=3$ components remain massless. This breaks $SU(2)$ to $U(1)$. 

Integrating out the massive fields, one can then in principle compute the low-energy
effective action of the massless fields. Restricting oneself to two-derivative terms,
the form of this action is constrained by $N=2$ susy to depend only on a single
holomorphic function $\F(a)$. Combining the knowledge of the one-loop computation of
$\F$ (the only perturbative contribution) and arguments about monodromies and the
expected number of singularities on moduli space, Seiberg and Witten were able to
determine $a(u)$ and $a_D(u)\equiv (\partial \F /\partial a)(u)$ explicitly in terms
of integrals of the holomorphic one-form on a certain genus one elliptic curve over
the cycles of the homology basis. From there, one can reconstruct $\F(a)$ and the
low-energy effective action.

The $a(u)$ and $a_D(u)$ however directly give the masses of the BPS states as
\begin{equation}
 m=\sqrt{2}\vert   n_e a(u) - n_m a_D(u)\vert
\label{i}
\end{equation}
where $n_e$ and $n_m$ are the integer electric and magnetic charge quantum numbers of
the state. The massive perturbative states (W bosons and superpartners) have $n_e=\pm
1,\  n_m=0$ so that $m=\sqrt{2} \vert a \vert$ as noted above. Magnetic monopoles that
do exist as non-pertubative solitonic states have $n_e=0, \ n_m=\pm 1$ and mass
$m=\sqrt{2} \vert a_D \vert$. States with masses strictly larger than the r.h.s. of
(\ref{i}) are non BPS states and must be in a different type of susy multiplet,  a
so-called long multiplet containing necessarily spins larger than one. All known
states turn out to be in short multiplets of spins less or equal to one and are BPS
states satifying (\ref{i}).

The functions $a(u)$ and $a_D(u)$ are explicitly given by
\begin{eqnarray}
a(u)&=& \left({u+1\over 2}\right)^{\scriptstyle 1\over \scriptstyle 2}\,
F\left(-{1\over 2}\raise 2pt\hbox{,}{1\over 2}\raise 2pt\hbox{,}1;{2\over u+1}\right)
\ ,   \cr
a_D(u)&=&  i\, {u-1\over 2}\, 
F\left({1\over 2}\raise 2pt\hbox{,}{1\over 2}\raise 2pt\hbox{,}2;{1-u\over 2}\right) \ .
\label{ii}
\end{eqnarray}
They have cuts and branch points. The latter are the singular points on the moduli
space at $u=1,\, -1$ and $\infty$. Around these points the section $(a_D,a)$ has
non-trivial monodromies in $SL(2,\Z)\equiv Sp(2,\Z)$. The full $Sp(2,\Z)$ is the group
of duality symmetries of the low-energy effective action and it includes
electric-magnetic duality. Note that the spectrum of (massive) BPS states will not be
invariant under the full duality group, since the latter is not a symmetry of the
non-abelian $N=2$ susy Yang-Mills theory, only of the abelian low-energy effective
action.

The singular points on moduli space have an important physical interpretation. There,
an otherwise massive BPS state becomes massless. While integrating out massive states
is a sensible operation, integrating out a massless field leads to divergencies. So
when a massive state that had been integrated out becomes massless somewhere on moduli
space, one expects and indeed gets a singularity at this point. 
The singularity at $u=+1$
corresponds to a magnetic monopole (as well as anti-monopole)  $(n_e,n_m)=\pm (0,1)$
becoming massless and the singularity at $u=-1$ to a dyon $(\pm 1,1)$ (and antidyon
$-(\pm 1,1)$) becoming massless (cf. eqs. (\ref{i}) and (\ref{ii})).

Note already that the two singularities are related by a $\Z_2$ symmetry acting on the
moduli space as $u\to -u$. Note also that as $u\to \infty$, $a(u) \sim \sqrt{2u} \to
\infty$. Since $a$ sets the mass scale, the $SU(2)$ Yang-Mills theory becomes
asymptotically free in this limit, which hence is the semi-classical limit. On the
other hand, $u=\pm 1$ are in a region of strong coupling and the associated
singularities are genuine non-perturbative effects.

\section{BPS STATES, CURVE OF MAR-\break GINAL STABILITY AND EXAMPLES OF DECAYS}

Equation (\ref{i}) gives the mass of a BPS state in terms of the corresponding central
charge
\begin{equation}
Z=n_e a(u) - n_m a_D(u)\ , \quad n_e,\ n_m \in \Z
\label{iii}
\end{equation}
appearing in the $N=2$ susy algebra. Note that $Z$ is given by the standard symplectic
invariant $\eta(p,\Omega)$ of $p=(n_e, n_m)$ and $\Omega=(a_D,a)$ which is such that 
$\eta(G p,G \Omega)=\eta(p,\Omega)$ for any $G\in Sp(2,\Z)$. Generically, for a given
point $u$ in moduli space, $a$ and $a_D$ are two complex numbers such that $a_D/a\notin
\R$ and all possible central charges form a lattice in the complex plane, see Fig. 1.
\begin{figure}[htb]
\vspace{9pt}
\centerline{ \fig{4.5cm}{4.5cm}{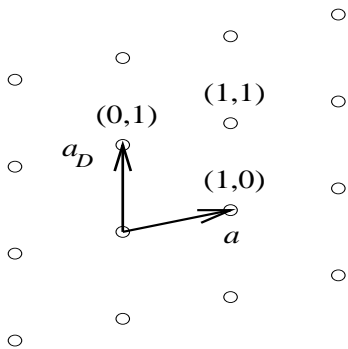} }
\caption{The lattice of central charges for generic $a_D$ and $a$}
\label{lattice}
\end{figure}
Each lattice point corresponds to an a priori possible BPS state 
$(n_e, n_m)$ whose mass is simply
its euclidean distance from the origin. Consider e.g. the dyon state
$(n_e,n_m)=(1,1)$. By charge conservation alone, it could decay into the W boson
$(1,0)$ and the magnetic monopole $(0,1)$ but, by the triangle inequality of elementary
geometry, the sum of the masses of the decay products would be larger than the mass of
the $(1,1)$ dyon. Hence the latter is stable. The same argument applies to all BPS
states $(n_e, n_m)$ such that $(n_e, n_m)\ne q(n,m)$ with $n,m,q\in\Z$, $q\ne \pm 1$:
states with $n_e$ and $n_m$ relatively prime are stable.

The preceeding argument fails if $(a_D/a)(u)\in \R$, since then the lattice collapses
onto a single line and decays of otherwise stable BPS states become possible. It is
thus of interest to determine  the set of all such $u$, i.e.
${\cal C}=\{u \in {\bf C}\  \vert\  (a_D(u)/a(u))\in \R\}$, which is called the curve 
of marginal stability \cite{ARG,MAT}. Given the explicit form of 
$a_D(u)$ and $a(u)$ it is
straightforward to determine \C\ numerically \cite{FB}, see Fig. 2, although it can
also be done analytically \cite{MAT}. 
\begin{figure}[htb]
\vspace{9pt}
\centerline{ \fig{6cm}{4cm}{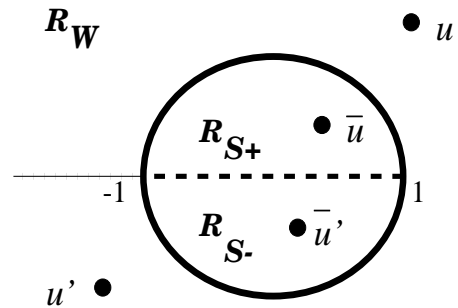} }
\caption{In the $u$ plane, we show the curve \C\ of marginal stability 
which is almost an ellipse centered
at the origin (thick line), 
the cuts of $a(u)$ and $a_D(u)$ (dotted and dashed lines), as well as the definitions of
the weak-coupling region $\rw$ and the
strong-coupling region ($\rsp \cup \rsm$).}
\label{curve}
\end{figure}
The precise form of the curve however is irrelevant for our
purposes. What is important is that as $u$ varies along the curve, $a_D/a$ takes all
values in $[-1,1]$. More precisely, if we call ${\cal C}^\pm$ the parts of \C\ in the
upper and lower half $u$ plane, then
\begin{eqnarray}
{a_D\over a}(u) &\in& [-1,0] \quad {\rm for}\ u\in {\cal C}^+ \ , \cr
&{}& \cr
{a_D\over a}(u) &\in& [0,1] \quad {\rm for}\ u\in {\cal C}^- \ ,
\label{iv}
\end{eqnarray}
with the value being discontinuous at $u=-1$ due to the cuts of $a_D$ and $a$ running
along the real axis from $-\infty$ to $+1$.

The curve \C\ separates the moduli space into two distinct regions: inside the curve
and outside the curve, see Fig. 2. If two points $u$ and $u'$ are in the same region,
i.e. if they can be joined by a path not crossing \C\ then the spectrum of BPS states
(by which we mean the set of quantum numbers $(n_e, n_m)$ that do exist) is
necessarily the same at $u$ and $u'$. Indeed, start with a given stable BPS state at
$u$. Then imagine deforming the theory adiabatically so that the scalar field $\f$
slowly changes its vacuum expectation value and $\langle {\rm tr \, }\f^2\rangle$
moves from $u$ to $u'$. In doing so, the BPS state will remain stable and it cannot
decay at any point on the path. Hence it will also exist at $u'$. If, however, $u$ and
$\tilde u$ are in different regions so that the path joining them must cross the curve
\C\ somewhere, then the initial BPS state will no longer be stable as one crosses the
curve and it can decay. Hence the spectrum at $u$ and $\tilde u$ need not be the same.

As an example, consider the possible decay of the W boson $(1,0)$ when crossing the
curve on ${\cal C}^+$ at a point where $a_D/a =r$ with $r$ any real number between
$-1$ and $0$. Charge conservation alone allows for the reaction
\begin{equation}
(1,0)\to (1,-1) + (0,1) \ .
\label{v}
\end{equation}
On ${\cal C}^+$, and only on ${\cal C}^+$, we also have the equality of masses, thanks
to
\begin{eqnarray}
\vert a+a_D\vert + \vert a_D\vert &=&\vert a \vert \left( \vert 1+r\vert +\vert r \vert
\right)  \cr
&=& \vert a \vert \left( 1+r -r\right) = \vert a \vert  \ .
\label{vi}
\end{eqnarray}
Had one crossed the curve in the lower half plane instead, $r$ would have been between
$0$ and $+1$ and the dyon $(1,-1)$ would have been 
decribed as $(1,1)$ (see below), and eq.
(\ref{vi}) would have worked out correspondingly.

Since the region of moduli space outside the curve contains the semi-classical domain
$u\to\infty$, we refer to this region as the semi-classical or weak-coupling region
${\cal R}_W$ and to the region inside the curve as the strong-coupling region
${\cal R}_S$. We call the
corresponding spectra also weak and strong-coupling spectra $\sw$ and ${\cal S}_S$.
This terminology is used due to the above-explained continuity of the spectra
throughout each of the two regions. Nevertheless, the physics close to the curve is
always strongly coupled even in the so-called weak-coupling region.

\section{THE MAIN ARGUMENT AND THE WEAK-COUPLING SPECTRUM}

The important property of the curve \C\ of marginal stability is

\noindent {\bf P1 : Massless states can only occur on the curve \C.}

\noindent
The proof is trivial: If we have a massless state at some point $u$, it necessarily is
a BPS state, hence $m(u) = 0$ implies $n_e a(u) - n_m a_D(u) = 0$ which can be
rewritten as $(a_D/a)(u) = n_e/n_m$. But $n_e/n_m$ is a real number, hence 
$(a_D/a)(u)$ is real, and thus $u\in$ \C. Indeed the points $u=\pm 1$ where the
magnetic monopole and the dyon $(\pm 1,1)$ become massless are on the curve. The
converse statement obviously also is true:

\noindent {\bf P2 : A BPS state $(n_e, n_m)$ with $n_e/n_m \in [-1,1]$ becomes
massless somewhere on the curve \C.}

\noindent
Of course, it will become massless precisely at the point $u\in$ \C\ where 
$(a_D/a)(u) = n_e/n_m$. Strictly speaking, in its simple form, this only applies to
BPS states in the weak-coupling region, since the description of BPS states in the
strong-coupling region is slightly more involved as shown below. Let me now state the
main hypothesis.

\noindent {\bf H : A state becoming massless always leads to a singularity of the
low-energy effective action, and hence of $a_D(u),\ a(u)$. The Seiberg-Witten solution 
(\ref{ii}) for $a_D(u),\ a(u)$ is correct and there are only two singularities at
finite $u$, namely $u=\pm 1$.}

\noindent
Then the argument we will repeatedly use goes like this: If a certain state would
become massless at some point $u$ on moduli space, it would lead to an extra
singularity which we know cannot exist. Hence this state either is the magnetic
monopole $\pm (0,1)$ or the $\pm (\pm 1,1)$ dyon and $u=\pm 1$, or this state cannot
exist.

As an immediate consequence we can show that the weak-coupling spectrum cannot
contain BPS states with $\vert n_m\vert > \vert n_e\vert >0$. Indeed, for such a state,
$n_e/n_m \in [-1,1]$ and it would be massless at the point $u$ on \C\ where 
$(a_D/a)(u) = n_e/n_m$. Since  $\vert n_m\vert > \vert n_e\vert >0$ it is neither the
monopole ($n_e=0$) nor the $(\pm 1,1)$ dyon, hence it cannot exist.

To determine which states are in $\sw$ one uses a global symmetry. Taking $u\to
e^{2\pi i} u$ along a path outside \C\ does not change the theory since one comes back
to the same point of moduli space, and hence must leave $\sw$ invariant. But it
induces a monodromy transformation
\begin{equation}
\pmatrix{ n_e\cr n_m\cr} \, \to \, M_\infty \pmatrix{ n_e\cr n_m\cr} \, , \,
M_\infty=\pmatrix{ -1&\hfill 2\cr \hfill 0 & -1\cr} .
\label{vii}
\end{equation}
In other words, $M_\infty \sw = \sw$. Now, we know that $\sw$ contains at least the
two states that are responsible for the singularities, namely $(0,1)$ and $(1,1)$
together with their antiparticles $(0,-1)$ and $(-1,-1)$. Applying $M_\infty^{\pm 1}$ 
on these
two states generates all dyons $(n,\pm 1),\ n\in \Z$. 
This was already clear from \cite{SW}. But
now we can just as easily show that there are no other dyons in the weak-coupling
spectrum. If there were such a state $\pm (k,m)$ with $\vert m\vert \ge 2$, then applying
$M_\infty^n,\ n\in \Z$, there would also be all states $\pm (k-2n m,m)$. The latter
would become massless somewhere on \C\ if $(k-2n m/m)=(k/m)-2n \in [-1,1]$. Since there
is always such an $n\in \Z$, this state, and hence $\pm (k,m)$ cannot exist in $\sw$.
Finally. the W boson which is part of the perturbative spectrum is left invariant by
$M_\infty$: $M_\infty (1,0) = - (1,0)$, where the minus sign simply corresponds to the
antiparticle. Hence we conclude
\begin{equation}
\sw = \left\{ \pm (1,0),\  \pm (n,1),\ n\in \Z \right\} \ .
\label{viii}
\end{equation}
This result was already known from semi-classical considerations on the moduli space
of multi-monopole configurations \cite{SEN,STERN}, but it is nice to rederive it in
this particularly simple way. Now let us turn to the new results of
\cite{FB} concerning the
strong-coupling spectrum.

\section{THE $\Z_2$ SYMMETRY}

The classical susy $SU(2)$ Yang-Mills theory has a $U(1)_R$ $R$-symmetry acting on the
scalar $\f$ as $\f\to e^{2 i \alpha}\f$ so that $\f$ has charge two. In the quantum
theory this global symmetry is anomalous, and it is easy to see from the explicit form
of the one-loop and instanton contributions to the low-energy effective action (i.e.
to $\F$) that only a discrete subgroup $\Z_8$ survives, corresponding to phases
$\alpha={2\pi \over 8} k,\ k\in\Z$. Hence under this $\Z_8$ one has $\f^2\to (-)^k
\f^2$. This $\Z_8$ is a symmetry of the quantum action and of the Hamiltonian, but a
given vacuum with $u=\langle {\rm tr \, }\f^2\rangle \ne 0$ is invariant only under
the $\Z_4$ subgroup corresponding to even $k$. The quotient (odd $k$) is a $\Z_2$
acting as $u\to -u$. Although a given vacuum breaks the full $\Z_8$ symmetry, the
broken symmetry (the $\Z_2$) relates physically equivalent but distinct vacua. In
particular, the mass spectra at $u$ and at $-u$ must be the same. This means that for
every BPS state $(n_e, n_m)$ that exists at $u$ there must be some BPS state $(\tilde
n_e, \tilde n_m)$ at $-u$ having the same mass:
\begin{equation}
\vert \tilde n_e a(-u) - \tilde n_m a_D(-u) \vert 
= \vert n_e a(u) - n_m a_D(u) \vert \ .
\label{ix}
\end{equation}
This equality shows that there must exist a matrix
$G\in Sp(2,\Z)$ 
such that
\begin{eqnarray}
\pmatrix{ \tilde n_e \cr \tilde n_m \cr}
&=&\pm G \pmatrix{ n_e\cr n_m\cr} \ , \cr
&{}& \cr
\pmatrix{ a_D\cr a\cr} (-u) &=& e^{i\omega}\,  G \pmatrix{ a_D\cr a\cr} (u)
\label{x}
\end{eqnarray}
where $e^{i\omega}$ is some phase. Indeed, from the explicit expressions (\ref{ii})
of $a_D$ and 
$a$ one finds, using standard relations between hypergeometric functions, that
\begin{equation}
G= G_{W,\epsilon} \equiv \pmatrix{ 1&\epsilon\cr 0& 1\cr}\, ,\ e^{i\omega}=
e^{-i\pi\epsilon/2}
\label{xi}
\end{equation}
where $\epsilon =\pm 1$ according to whether $u$ is in the upper or lower half plane.
The subscript $W$ indicates that this is the matrix to be used in the weak-coupling
region, while for the strong-coupling region there is a slight subtlety to be
discussed soon. We have just shown that  for any BPS state $(n_e, n_m)$ existing at
$u$ (in the weak-coupling region) with mass $m$ there exists another BPS state
$(\tilde n_e, \tilde n_m)=\pm G_{W,\epsilon} (n_e, n_m)$ at $-u$ with the same mass
$m$. Now, since both $u$ and $-u$ are outside the curve \C, they can be joined by a
path never crossing \C, and hence the BPS state $(\tilde n_e, \tilde n_m)$ must also
exist at $u$, although with a different mass $\tilde m$. So we have been able to use
the broken symmetry to infer the existence of the state $(\tilde n_e, \tilde n_m)$ at
$u$ from the existence of  $(n_e, n_m)$ at the {\it same} point $u$ of moduli space.
Starting from the magnetic monopole $(0,1)$ at $u$ in the upper half plane (outside
\C) one deduces the existence of all dyons $(n,1)$ with $n\ge 0$. Taking similarly $u$
in the lower half plane (again outside \C) one gets all dyons $(n,1)$ with $n\le 0$.
The W boson $(1,0)$ is invariant under $G_{W,\epsilon}$. Once again, one generates
exactly the weak-coupling spectrum $\sw$ of (\ref{viii}), and clearly $G_{W,\epsilon}
\sw = \sw$.

\section{THE STRONG-COUPLING SPECTRUM}

It is in the strong-coupling region that this $\Z_2$ symmetry will show its full
power. Here $M_\infty$ no longer is a symmetry, since a monodromy circuit around
infinity can be deformed all through the weak-coupling region but it cannot cross \C\
into the strong-coupling region since the state that is taken along this circuit may
well decay upon crossing the curve \C. The relations (\ref{x},\ref{xi}) 
expressing $a_D(-u),
a(-u)$ in terms of $a_D(u), a(u)$ nevertheless remain true. What needs to be
reexamined is the relation between $\tilde n_e, \tilde n_m$ and $n_e, n_m$. This is
due to the fact that there is a cut of the function $a(u)$ running between $-1$ and
$1$, separating the strong-coupling region ${\cal R}_S$ into two parts, 
$\rsp$ and $\rsm$, 
as shown in Fig. 2.
As a consequence, the same BPS state is described by two different sets of
integers in $\rsp$ and $\rsm$. If we call the corresponding spectra $\ssp$ and $\ssm$
then we have
\begin{eqnarray}
\ssm&=&M_1^{-1} \ssp\ ,\ \pmatrix{n_e'\cr n_m'\cr} = M_1^{-1} \pmatrix{n_e\cr n_m\cr}
\ ,\cr
M_1^{-1}&=&\pmatrix{1&0\cr 2&1\cr} \ .
\label{xia}
\end{eqnarray}
This change of description is easily explained: take a BPS state $(n_e, n_m)\in \ssp$
at a point $u\in \rsp$ and transport it to a point $u'\in \rsm$. In doing so, its
mass varies continuously and nothing dramatic can happen since one does not cross the
curve \C. Hence, as one crosses from $\rsp$ into $\rsm$, 
the functions $a_D$ and $a$ must also vary
smoothly, which means that at $u'\in\rsm$ one has the analytic continuation of
$a_D(u)$ and $a(u)$. But this is not what one calls $a_D$ and $a$ in $\rsm$. Rather,
these analytic continuations $\tilde a_D(u')$ and $\tilde a(u')$ are related to
$a_D(u')$ and $a(u')$ by the monodromy matrix around $u=1$ which is $M_1$ as
\begin{equation}
\pmatrix{\widetilde {a_D}(u')\cr \tilde a(u')\cr} = 
M_1 \pmatrix{ a_D(u')\cr a(u')\cr} \ .
\label{xib}
\end{equation}
Hence the
mass of the BPS state at $u'$ is $\sqrt{2} \vert n_e \tilde a(u') - n_m \tilde
a_D(u')\vert$ $= \sqrt{2}\vert n_e' a(u') - n_m' a_D(u') \vert$ where $n_e',\, n_m'$
are given by eq. (\ref{xia}). As a consequence of the two different descriptions of
the same BPS state, the $G$-matrix implementing the $\Z_2$ transformation on the
spectrum has to be modified. As before, from 
the existence of $(n_e, n_m)$ at $u\in \rsp$ one
concludes the existence of a state $G_{W,+} (n_e, n_m)$ at $-u\in \rsm$. This
same state must then also exist at $u$ but is described as 
$M_1 G_{W,+} (n_e, n_m)$. Had one started with a $u\in \rsm$ the relevant matrix would
have been $M_1^{-1} G_{W,-}$. Hence, in the strong-coupling region $G_{W,\pm}$ is
replaced by
\begin{equation}
G_{S,\epsilon}=(M_1)^\epsilon G_{W,\epsilon} = \pmatrix{\hfill 1&\hfill\epsilon\cr
-2\epsilon&-1\cr}\ ,
\label{xii}
\end{equation}
and again one concludes that the existence of a BPS state $(n_e, n_m)$ at $u\in {\cal
R}_{S,\epsilon}$ implies the existence of another BPS state $G_{S,\epsilon} (n_e,
n_m)$ at the {\it same} point $u$. The important difference now is that
$G_{S,\epsilon}^2=-{\bf 1}$, so that applying this argument twice just gives back
$(-n_e, -n_m)$. But this is the antiparticle of $(n_e, n_m)$ and always exists
together with $(n_e, n_m)$. As far as the determination of the spectrum is concerned
we do not really need to distinguish particles and antiparticles. In this sense,
applying $G_{S,\epsilon}$ twice gives back the same BPS state. Hence in the
strong-coupling region, all BPS states come in pairs, or $\Z_2$ doublets (or quartets
if one counts particles and antiparticles separately):
\newpage
\begin{eqnarray}
\pm \pmatrix{n_e\cr n_m\cr}\in \ssp \, &\Leftrightarrow& \, \pm\,  G_{S,+} 
\pmatrix{n_e\cr n_m\cr} \cr
&=&\pm \pmatrix{ n_e+n_m\cr -2n_e-n_m\cr} \in \ssp \cr
&{}& 
\label{xiii}
\end{eqnarray}
and similarly for $\ssm$.
An example of such a doublet is the magnetic monopole $(0,1)$ and the dyon
$(1,-1)=-(-1,1)$ which are the two states becoming massless at the $\Z_2$-related
points $u=1$ and $u=-1$. Note that in $\ssm$ the monopole is still described as
$(0,1)$ while the same dyon is described as $(1,1)$. It is now easy to show that this
is the only doublet one can have in the strong-coupling spectrum. Indeed, one readily
sees that either $n_e/n_m \equiv r$ is in $[-1,0]$ or $(n_e+n_m)/(-2 n_e-n_m)= -
(r+1)/(2r+1)$ is in $[-1,0]$. This means that one or the other member of the $\Z_2$
doublet (\ref{xiii}) becomes massless somewhere on ${\cal C}^+$, the part of the curve
\C\ that can be reached from $\rsp$. But as already repeatedly argued, the only states
ever becoming massless are the magnetic monopole $(0,1)$ and the dyon $(1,-1)$. Hence
no other $\Z_2$ doublet can exist in the strong-coupling spectrum and we conclude that
\begin{eqnarray}
\ssp &=& \left\{ \pm (0,1), \pm (-1,1) \right\} \cr
&\Leftrightarrow& \cr
\ssm &=& \left\{ \pm (0,1), \pm (1,1) \right\} \ .
\label{xiv}
\end{eqnarray}
{\bf P3 : The strong-coupling spectrum consists of only those BPS states that are
responsible for the singularities. All other weak-coupling, i.e. semi-classical BPS
states must and do decay consistently into them when crossing the curve \C.}

\noindent
We have shown above the example of the decay of the 
W boson, cf. eq. (\ref{v}), but it is just as
simple to show consistency of the other decays \cite{FB}.

When adding massless quark hypermultiplets next, we will see that the details of the
spectrum change, however, the conclusion P3 will remain the same.

\section{GENERALISATION TO $N=2$ SUSY QCD INCLUDING $N_f=1,2,3$ 
MASSLESS QUARK HYPERMULTIPLETS}

We will continue to consider only the gauge group SU(2) as studied in \cite{SWII}. We
will also restrict ourselves to the case of vanishing bare masses of the quark
hypermultiplets. Here, we will be very qualitative and describe only the results,
referring the reader to \cite{BF} for details. The main difference with respect to the
previous case of pure Yang-Mills theory is that now the BPS states carry
representations of the flavour group which is the covering group of $SO(2N_f)$, namely
$SO(2)$ for one flavour, $Spin(4)=SU(2)\times SU(2)$ for two flavours, and
$Spin(6)=SU(4)$ for three flavours. We will present each of the three cases separately.

\subsection{$N_f=1$}

According to Seiberg and Witten \cite{SWII} there are 3 singularities at finite points
of the Coulomb branch of the moduli space. They are related by a global discrete $\Z_3$
symmetry. This $\Z_3$ is the analogue of the $\Z_2$ symmetry discussed previously. Its
origin is slightly more complicated, however, since the original $\Z_{12}$ is due to a
combination of a $\Z_6$ coming from the anomalous $U(1)_R$ symmetry and of the
anomalous flavour-parity of the $O(2N_f)$ flavour group. In any case, the global
discrete symmetry of the quantum theory is $\Z_{12}$. The vacuum with non-vanishing
value of $u=\langle {\rm tr\, }\f^2 \rangle$ breaks this to $\Z_4$. The quotient $\Z_3$
acting as $u\to e^{\pm 2\pi i/3} u$ then is a symmetry relating different but
physically equivalent vacua. The three singular points are due to a massless
monopole $(0,1)$, a massless dyon $(-1,1)$ and another massless dyon $(-2,1)$. Again
there is a curve of marginal stability that was obtained from the explicit expressions
for $a_D(u)$ and $a(u)$ \cite{BF}. It is almost a circle, and of course, it goes
through the three singular points, see Fig. 3, 
\begin{figure}[htb]
\vspace{9pt}
\centerline{ \fig{5.5cm}{4.5cm}{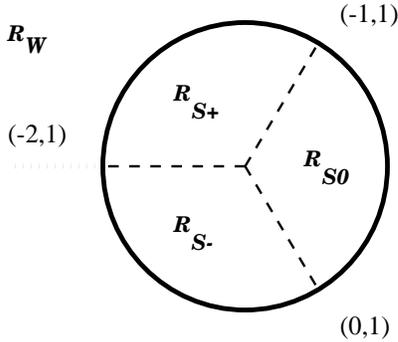} }
\caption{The curve of marginal stability and the three different portions of the
strong-coupling region separated by the cuts, for $N_f=1$}
\label{matter}
\end{figure}
where we also indicated the various cuts
and correspondingly different portions $\rsp, \rsm, \rso$ of the
strong-coupling region ${\cal R}_S$. So here one needs to introduce three different
desciptions of the same strong-coupling BPS state.
The corresponding spectra are denoted ${\cal S}_{S+}$,
${\cal S}_{S0}$ and ${\cal S}_{S-}$. The ratio $a_D/a$ increases
monotonically from $-2$ to $+1$ as one goes along the curve in a clockwise sense,
starting at the point where $(-2,1)$ is massless. Then using exactly the same type of
arguments as we did before, one obtains the weak and strong-coupling spectra. All
states in the latter now belong to a single $\Z_3$ triplet, containing precisely the
three states responsible for the singularities. Denoting a BPS state by $(n_e, n_m)_S$
where $S$ is the $SO(2)$ flavour charge, and denoting its antiparticle $(-n_e,
-n_m)_{-S}$ simply by $-(n_e, n_m)_S$, one finds \cite{BF}
\begin{eqnarray}
\sw &=& \left\{ \pm (2,0)_0 , \ \pm (1,0)_1 ,\  \pm (2n,1)_{1/2} , \right.\cr
&{}& \phantom{\{ } \left. \pm (2n+1,1)_{-1/2} ,\ n\in \Z \right\} \cr
&{}& \cr
\sso &=&  \left\{ \pm (0,1)_{1/2} , \ \pm (-1,1)_{-1/2} , \ \pm (1,0)_{1/2} \right\} \cr
&{}&
\label{xv}
\end{eqnarray}
with states in $\ssp$ or $\ssm$ related to the description in $\sso$ by the appropriate
monodromy matrices: $\ssp = \pmatrix{\hfill 2&1\cr -1 & 0\cr} \sso$, $\ssm = 
\pmatrix{1&0\cr 1&1\cr} \sso$. One sees that the state $(1,0)_{1/2}$ in $\sso$
corresponds to $(2,-1)_{1/2}$ in $\ssp$ or to $(1,1)_{1/2}$ in $\ssm$ and is the one
responsible for the third singularity. Also note that  following \cite{SWII,BF} we
changed the normalisation of the electric charge  by a factor of 2, so that now
$(2,0)_0$ is the W boson and $(1,0)_1$ is the quark. All decays across the curve \C\
are consistent with conservation of the mass and of all quantum numbers, i.e. electric
and magnetic charges, as well as the $SO(2)$ flavour charge. For example, when crossing
\C\ into ${\cal R}_{S0}$, the quark decays as $(1,0)_1 \to (0,1)_{1/2}+(1,-1)_{1/2}$.

\subsection{$N_f=2$}

This case is very similar to the pure Yang-Mills case. The global discrete symmetry
 acting on the Coulomb branch of moduli space is again $\Z_2$ and the curve of marginal
stability is exactly the same, cf. Fig. 2, with the singularities again due to a
massless magnetic monopole $(0,1)$ and a massless dyon $(1,1)$. Note however, that this
is in the new normalisation where the W boson is $(2,0)$. So this dyon has half the
electric charge of the W, contrary to what happened for $N_f=0$. With the present
normalisation one finds the weak and strong-coupling spectra as
\begin{eqnarray}
\sw &=& \left\{ \pm (2,0) , \ \pm (1,0) ,\  \pm (n,1) ,\ n\in \Z \right\} \cr
&{}& \cr
\ssp &=&  \left\{ \pm (0,1) , \ \pm (-1,1) \right\}
\label{xvi}
\end{eqnarray}
and all decays across \C\ are again consistent with all quantum numbers. For the quark
one has e.g. $(1,0)\to (0,1)+(1,-1)$ with the flavour representations of $SU(2)\times
SU(2)$ working out as $({\bf 2}, {\bf 2}) = ({\bf 2}, {\bf 1})\otimes ({\bf 1}, {\bf
2})$.

\subsection{$N_f=3$}

In this case the global symmetry of the action is $\Z_4$ and a given vacuum is
invariant under the full $\Z_4$. Consequently, there is no global discrete symmetry
acting on the Coulomb branch of the moduli space. There are two singularities
\cite{SWII}, one due to a massless monopole, the other due to a massless dyon $(-1,2)$
of {\it magnetic} charge 2. The existence of magnetic charges larger than 1 is a
novelty of $N_f=3$. 
The curve of marginal stability again goes through the two singular
points. It is a shifted and rescaled version of the corresponding curve for $N_f=0$,
see Fig. 2. Due to the cuts, again we need to introduce two different descriptions
of the same strong-coupling BPS state. The variation of $a_D/a$ along the curve \C\ is
from $-1$ to $-1/2$ on ${\cal C}^+$ and from $-1/2$ to $0$ on ${\cal C}^-$. Luckily,
this is such that we do not need any global symmetry to determine the strong-coupling
spectrum. For the weak-coupling spectrum, one uses the $M_\infty$ symmetry. One finds
\begin{eqnarray}
\sw &=& \left\{ \pm (2,0) , \ \pm (1,0) ,\  
\pm (n,1) ,\right.\cr
&{}& \phantom{ \{ } \left. \pm (2n+1,2) , \ n\in \Z \right\} \cr
&{}& \cr
\ssp &=&  \left\{ \pm (1,-1) , \ \pm (-1,2) \right\}
\label{xvii}
\end{eqnarray}
with $(1,-1)\in \ssp$ corresponding to $(0,1)\in \ssm$, so this is really the magnetic
monopole.

The flavour symmetry group is $SU(4)$, and the quark $(1,0)$ is in the representation
${\bf 6}$, the W boson $(2,0)$ and the dyons of magnetic charge two are singlets, while
the dyons $(n,1)$ of magnetic charge one  are in the representation 
${\bf 4}$ if $n$ is even and in
${\overline {\bf 4}}$ if $n$ is odd. Antiparticles are in the complex conjugate
representations of $SU(4)$. Again, all decays across the curve \C\ are consistent with
all quantum numbers, and in particular with the $SU(4)$ Clebsch-Gordan series. As an
example, consider again the decay of the quark, this time as $(1,0)\to 2\times (0,1) + (1,-2)$.
The representations on the l.h.s. and r.h.s. are ${\bf 6}$ and ${\bf 4}\otimes {\bf
4}\otimes {\bf 1}$. Since ${\bf 4}\otimes {\bf 4}={\bf 6} \oplus {\bf 10}$ this decay
is indeed consistent. All other examples can be found in \cite{BF}.

\section{CONCLUSIONS}

The conclusions have already been listed in the abstract, and there is no need to
repeat them here again.

\vskip 2.mm
\noindent{\bf Acknowledgements}

It is a pleasure to thank the organizing committee of the SUSY'96 conference at
College Park, MD, and in particular Jim Gates, for the invitation to present the
material covered in these notes.

\noindent{

\end{document}
\begin{thebibliography}{9}

\bibitem{SW} N. Seiberg and E. Witten, {\it Electric-magnetic duality, monopole
condensation, and confinement in $N=2$ supersymmetric Yang-Mills theory}, 
\NP B426 1994 19 , {\tt hep-th/9407087}.

\bibitem{CV} S. Cecotti, P. Fendley, K. Intriligator and C. Vafa,
{\it A new supersymmetric index}, \NP B386 1992 405 ;\hfill\break
S. Cecotti and C. Vafa, {\it On classification of $N=2$
supersymmetric theories}, \CMP 158 1993 569 .

\bibitem{FB} F. Ferrari and A. Bilal, {\it The strong-coupling spectrum of
Seiberg-Witten theory},  Nucl. Phys. {\bf B469} (1996) 387,
{\tt hep-th/9602082}.

\bibitem{BF} A. Bilal and F. Ferrari, {\it Curves of marginal stability, and weak
and strong-coupling BPS spectra in $N=2$ supersymmetric QCD}, \'Ecole Normale
Sup\'erieure preprint LPTENS-96/22, {\tt hep-th/9605101}.

\bibitem{ARG} P.C. Argyres, A.E. Faraggi and A.D. Shapere, {\it Curves
of marginal stability in $N=2$ super-QCD}, preprint IASSNS-HEP-94/103, 
UK-HEP/95-07, {\tt hep-th/9505190};\hfill\break
A. Fayyazuddin, {\it Some comments on $N=2$ supersymmetric
Yang-Mills}, \MPL A10 1995 2703 , {\tt hep-th/9504120}.

\bibitem{MAT} M. Matone, {\it Koebe $1/4$-theorem and inequalities in
$N=2$ super-QCD}, \PR D53 1996 7354 , {\tt hep-th/9506181}.

\bibitem{SEN} A. Sen, {\it Dyon-monopole bound states, self-dual harmonic
forms on the multi-monopole moduli space, and $SL(2,\Z)$ invariance
in string theory}, \PL B329 1994 217 , {\tt hep-th/9402032}.

\bibitem{STERN} S. Sethi, M. Stern and E. Zaslow, {\it Monopole and
dyon bound states in $N=2$ supersymmetric Yang-Mills theories},
\NP B457 1995 484 , {\tt hep-th/9508117}.

\bibitem{SWII}  N. Seiberg and E. Witten,  {\it Monopoles, duality and chiral
symmetry breaking in $N=2$ supersymmetric QCD}, \NP B431 1994 484 ,
{\tt hep-th/9408099}.


\end{thebibliography}
